\documentclass[10pt,english,fleqn,refstyle]{revtex4-1} 

\usepackage[T1]{fontenc}
\usepackage[latin9]{inputenc}
\usepackage{xcolor}
\usepackage{refstyle}
\usepackage{textcomp}
\usepackage{amsmath}
\usepackage{amssymb}
\usepackage{graphicx}
\usepackage{esint}
\PassOptionsToPackage{normalem}{ulem}
\usepackage{ulem}

\usepackage{amsmath,amsfonts,amssymb,amsthm}
\usepackage{hyperref}
\usepackage[fleqn]{mathtools}
\usepackage{enumitem}

\theoremstyle{plain}

\newtheorem*{theorem*}{Theorem}

\newcommand{\supp}{\mathrm{supp}}
\newcommand{\suml}{\sum\limits}
\newcommand{\intl}{\int\limits}
\newcommand{\liml}{\lim\limits}

\newcommand{\E}{\mathbf{E}}
\renewcommand{\H}{\mathbf{H}}
\newcommand{\D}{\mathbf{D}}
\newcommand{\B}{\mathbf{B}}

\newcommand{\Phib}{\mathbf{\Phi}}

\newcommand{\eps}{\varepsilon }
\renewcommand{\a}{\alpha }
\renewcommand{\b}{\beta }
\newcommand{\g}{\gamma }

\newcommand{\Loc}{\mathbf{L}^2_{\rm loc}}
\newcommand{\Locs}{\mathsf{L}^2_{\rm loc}}
\renewcommand{\L}{\mathbf{L}^2 }
\newcommand{\Ls}{\mathsf{L}^2 }
\newcommand{\C}{\mathbf{C}_0^\infty}
\newcommand{\Cs}{\mathsf{C}_0^\infty}
\newcommand{\Dp}{\mathbf{D}'}
\newcommand{\Dps}{\mathsf{D}'}

\newcommand{\Rp}{\mathbb{R}^3_+}
\newcommand{\Rm}{\mathbb{R}^3_-}
\newcommand{\R}{\mathbb{R}^3}

\newcommand{\TE}{\mathrm{TE}}
\newcommand{\TM}{\mathrm{TM}}

\newcommand{\curl}{\nabla\times}
\renewcommand{\div}{\nabla\cdot}

\newcommand{\n}{\mathbf{n}} 
\newcommand{\tn}{\times \n}    
\renewcommand{\r}{\mathbf{r} } 
\newcommand{\s}{\mathbf{s}}  
\newcommand{\Id}{\mathrm{I}} 
\renewcommand{\k}{\mathbf{k} } 
\renewcommand{\tt}{\mathrm{T}} 
\newcommand{\e}{\mathrm{e}} 
\renewcommand{\d}{\hspace{1mm}\mathrm{d}}
\newcommand{\ds}{\displaystyle}
\renewcommand{\i}{\mathrm{i}}

\allowdisplaybreaks

\begin{document}

\title[Interface conditions between vacuum and metamaterial]{Interface conditions for a metamaterial with strong spatial dispersion}

\author{Andrii Khrabustovskyi}
\affiliation{Institute for Analysis, Karlsruhe Institute
of Technology, Englerstra{\ss}e 2, 76131 Karlsruhe, Germany}

\author{Karim Mnasri}
\affiliation{Institute of Theoretical Solid State Physics, Karlsruhe Institute
of Technology, Wolfgang-Gaede-Str. 1, 76131 Karlsruhe, Germany }

\author{Michael Plum}
\affiliation{Institute for Analysis, Karlsruhe Institute
of Technology, Englerstra{\ss}e 2, \,   76131 Karlsruhe, Germany}

\author{Christian Stohrer}
\affiliation{Institute for Applied and Numerical Mathematics, Karlsruhe Institute
of Technology, Englerstra{\ss}e 2, 76131 Karlsruhe, Germany}

\author{Carsten Rockstuhl}
\affiliation{Institute of Theoretical Solid State Physics, Karlsruhe Institute
of Technology, Wolfgang-Gaede-Str. 1, 76131 Karlsruhe, Germany }
\affiliation{Institute of Nanotechnology, Karlsruhe Institute of Technology,
P.O. Box 3640, 76021 Karlsruhe, Germany}

\begin{abstract}
Local constitutive relations, i.e. a weak spatial dispersion, are usually considered in the effective description of metamaterials. However, they are often insufficient and effects due to a nonlocality, i.e. a strong spatial dispersion, are encountered. Recently (K.~Mnasri {\it et al.}, arXiv:1705.10969), a generic form for a nonlocal constitutive relation has been introduced that could accurately describe the bulk properties of a metamaterial in terms of a dispersion relation. However, the description of functional devices made from such nonlocal metamaterials also requires the identification of suitable interface conditions. In this contribution, we derive the interface conditions for such nonlocal metamaterials.
\end{abstract}
\maketitle

\section{Introduction}
Metamaterials greatly enhance our abilities to control the propagation of electromagnetic waves. They derive their properties from unit cells, called the meta-atoms, that interact with light in quite a peculiar way. Metamaterials attracted attention as meta-atoms can be perceived that show, e.g. a strong magnetic response. As the properties of the meta-atom translate to the properties of the entire material, novel materials came in reach that possess a magnetic dispersion. Such materials are of fundamental scientific interest and unlock simultaneously a plethora of applications. Exemplarily, we can mention perfect lenses \cite{GUENNEAU2009352}, cloaks \cite{Silveirinha2008}, broadband anti-reflection coatings \cite{Spinelli2012,Fahr2013}, or directional antennas \cite{Alu2007}; but there exist many more examples. 

A key requirement to link the actual metamaterial made from structured unit cells to a conceptual homogenous material is the assignment of effective properties. Of course, describing the light propagation through a device that has metamaterial constituents can always be done while solving the full Maxwell problem, but this is impractical for larger structures and indeed quite tedious. Moreover, just as we do not consider ordinary materials in the context of electrodynamics as being composed of atoms, it is the philosophy of an actual metamaterial to discuss it in terms of emerging properties, i.e. effective material properties. Then, the metamaterial is considered as homogenous.

The procedure of assigning effective properties to a metamaterial is performed within a parameter retrieval. The effective parameters then shall allow to describe the light propagation through the metamaterial with just the same accuracy as a full wave solution would do. Technically, there are different ways of assigning effective properties \cite{0953-8984-21-15-155404,PhysRevE.85.066603,Sihvola2013364,Chipouline201277,PhysRevB.88.125131,PhysRevB.93.024418,1367-2630-15-11-113044,Shevchenko:17,PhysRevB.87.235136}. Basic averaging procedures \cite{Chipouline201277} that consider the metamaterial as an ensemble of polarisable dipoles do the job if the metamaterial is strongly diluted and the polarizability of the meta-atoms is not too strong. For most metamaterials made of strongly scattering meta-atoms that are densely packed, retrieval procedures were developed that rely on considering reflection and transmission from a slab \cite{Grahn:13,PhysRevB.77.195328,Liu20130240,PhysRevB.89.245102,6529320} or that average the fields in the unit cells \cite{Chipouline201277, Smith:06}. Also, retrievals based on the dispersion relation and the wave impedance of the Bloch modes were suggested \cite{Lawrence:13,PhysRevB.86.035127,Paulimpedance,RockstuhlFishnet} or by dual field interpolation in Ref.~\onlinecite{PhysRevE.84.016609}. However, how such a retrieval is done is a rather technical question. At the heart, the philosophy of all these methods is to postulate a specific constitutive relation the metamaterial shall obey and then to link the parameters of these constitutive relations to parameters that can be measured or simulated. 

In all these aforementioned retrieval methods, local constitutive relations were considered. Under such assumption, the electric field $\vec{\rm E}(\r,t)$ and the magnetizing field $\vec{\rm H}(\r,t)$ are linked locally to the electric displacement field $\vec{\rm D}(\r,t)$ and the magnetic field $\vec{\rm B}(\r,t)$. It has been shown that to observe a local magnetic response for a given metamaterial it has to have actually an electric response that corresponds to weak spatial dispersion \cite{PhysRevB.75.115104,PhysRevB.78.165114,MenzelValidity}. The weak spatial dispersion is captured by a constitutive relation where the electric displacement depends on the second spatial derivative of the electric field at the considered point in space. In frequency space it reads as 
\begin{gather}\label{basic}
D_i(\r,\omega)= \epsilon_{ij}(\omega)E_j(\r,\omega)+a_{ijkl}(\omega)\partial_k\partial_lE_j(\r,\omega).
\end{gather}
A possible first derivative is linked to bi-anisotropic effects and shall not be considered here, i.e. we consider central-symmetric metamaterials. Therefore, it is omitted. By exploiting a suitable gauge transformation, this weak spatial dispersion in the electric response can be recast to appear as a local magnetic response. This, however, is only possible if the following assumption holds
\begin{gather}\label{requirement}
a_{ijkl}(\omega)\partial_j\partial_kE_l(\r,\omega)\stackrel{!}{=} \left(\nabla\times\left[\a\nabla\times\E(\r,\omega)\right]\right)_i.
\end{gather}
This assumption is not valid for centro-symmetric systems but is a prerequisite to end up with local constitutive relations; hence it has been assumed to be valid. This is quite practical as for a local medium the interface conditions are known. These interface conditions, in general, express how the different fields involved behave at an interface from one medium to another. For a local medium we require that the tangential component of the electric and the magnetizing field and the normal component of the electric displacement and the magnetic field are continuous. Having interface conditions available is of fundamental importance to describe the light propagation through any kind of device that is made from a metamaterial. We need to know them as soon as the metamaterial is no longer infinitely extended in all spatial directions. 

However, it got more and more obvious that such a local description is quite limited for two reasons. Foremost, it is easy to verify that the assumption postulated in expression \ref{requirement} is not always justified. For a centro-symmetric metamaterial, a symmetry frequently encountered, additional terms of the form (cf. Ch. 3.2 in Ref~\onlinecite{Agranovich1984}) 
\begin{gather}\label{symmetry_extension}
\suml_{j\in \{x,y,z\}}{\partial_j}\left(a_{jjjj} {\partial_j\E}\right)
\end{gather}
appear that are strictly required to be zero if the local description shall be valid. However, this cannot be guaranteed {\it per se}. Moreover, there is clear evidence that it is not justified to truncate the Taylor expansion in the constitutive relation after the second term, which essentially leads to Eqn.~\ref{basic}. Whereas local constitutive relations strictly lead to elliptical or hyperbolic dispersion relations, the numerically computed dispersion relation of selected metamaterials possesses such properties only in quite a limited reciprocal space, i.e. for small values of $k_x$ and $k_y$ when considering the $z$-axis as the principle propagation direction. To accurately describe the dispersion relation of metamaterials requires to take higher order terms in the aforementioned Taylor expansion into account. As we concentrate on centro-symmetric materials where all odd orders vanish, the next higher order terms would be the fourth order term. {Nonlocal materials, in general, are characterized by constitutive relations where the response is a functional of a set of fields over the material or of derivatives of the field. Such nonlocal materials are sometimes also refereed to as nonsimple \cite{MabrizioMorro03}}.

Advanced constitutive relations that mitigate both problems were  suggested only recently \cite{2017arXiv170510969M}. They constitute an immediate extension to the constitutive relations considered in Eqn.~\ref{basic} and read as: 
\begin{gather}\label{const-relations_test}
\D(\r,\omega)= \eps(\omega)\E(\r,\omega)+\curl\left(\a(\omega)\curl\E(\r,\omega)\right)
+\suml_{j\in \{x,y,z\}}{\partial_j}\left(a_{jjjj} {\partial_j\E}\right)+
\curl\curl\left( \g(\omega)\curl\curl\E(\r,\omega)\right)
\end{gather}
where a tensorial notation has been introduced and the exact meaning of the quantities will be explained in detail below. Note that  for the last term an assumption comparable to the assumption in expression \ref{requirement} has been enforced. In particular, we require that
\begin{gather}\label{requirement_fourth}
\suml_{j,k,l,m,n}a_{ijklmn}(\omega)\partial_j\partial_k\partial_l\partial_mE_n(\r,\omega)\stackrel{!}{=} \left(\nabla\times\nabla\times\left[\a\nabla\times\nabla\times~\E(\r,\omega)\right]\right)_i
\end{gather}
holds. Only with such an assumption the dispersion relation will be of the same order as the dispersion relation that emerges when considering the term in expression \ref{symmetry_extension} as the extension. Therefore, they appear on equal footing. Please also note, in what follows we will use the notations $\beta_j$ instead of $a_{jjjj}$. It has to be stressed that also other nonlocal constitutive relations where explored in the context of metamaterials but they frequently assume a specific geometry \cite{PhysRevB.91.184207,PhysRevB.86.085146,PhysRevB.92.085107,Cabuz:11,Tsukerman:11}.  Also, additional boundary conditions have to be considered to solve the interface problems. These additional boundary conditions are usually introduced on phenomenological grounds \cite{AGRANOVICH1973121,1367-2630-12-11-113047} and were controversially discussed in Refs.~\onlinecite{PhysRevLett.80.2889,PhysRevLett.83.1263,PhysRevLett.83.1264}.

Recently \cite{2017arXiv170510969M}, we could show that the dispersion relation of a referential metamaterial can be much better described with the constitutive relation of Eqn.~\ref{const-relations_test}. However, this analysis only considered the bulk properties, and the characteristics of the eigenmodes were discussed. To fully capitalise on such constitutive relations, we need to equip them with suitable interface conditions to describe the optical response from basic functional elements made from such metamaterials. Potentially the simplest example for such functional element is a slab and we wish to know how light is reflected and transmitted from such a slab. 

In this contribution, we derive the urgently needed interface conditions from first principles by relying on a generalized (weak) formulation of Maxwell's equations. These interface conditions allow to find the Fresnel equations for the reflection and transmission of a plane wave and for both polarizations from a slab. These Fresnel equations are also derived in this contribution. The paper is made to be as self-sustained as possible, which explains its mathematical attitude. 

The manuscript is structured such that in Section~\ref{sec1} we detail Maxwell's equations, the constitutive relations we are looking at, and the considered geometry. In Section~\ref{sec2} we discuss their generalized solutions. In Section~\ref{sec3} we derive the associated interface conditions and in Section~\ref{sec4} we outline the Fresnel equations to compute the reflection and transmission from a slab of such a homogenous metamaterial characterised by nonlocal constitutive relations. The interface conditions we derive in Section~\ref{sec3} are valid under specific smoothness assumptions. However, they remain valid also without these smoothness assumptions in a generalised sense which requires the concept of traces. Therefore, the last section is devoted to this concept of the traces. We conclude on our work in a devoted finalizing section. Finally, in  an Appendix we present a short overview on the concept of generalized functions.

Prior starting with the derivation, we introduce some notations that will be used throughout the paper. Below, $\Omega$ is an \textit{open} domain in $\mathbb{R}^n$.

\begin{itemize} 

\item By $\r=(x,y,z)$ we denote points in $\R$ (spatial variable), by $t\in\mathbb{R}$ we denote a time variable.\smallskip

\item $\Rp:=\{\r\in\R:\ z>0\}$, $\Rm:=\{\r\in\R:\ z<0\}$, $\Gamma=\{\r\in\R:\ z=0\}.$ They denote the different half-spaces above and below the interface we consider.

\item  $\nabla$, $\div$, $\curl$ are, respectively, the gradient, divergence, and curl with respect to $\r$.\smallskip

\item $^{\tt}$ is the operation of transposition.\smallskip

\item $\n=(0,0,1)^{\tt}$ is the unit normal vector on $\Gamma$.\smallskip

\item For $\Psi:\Gamma\to\mathbb{C}$ we set $\nabla_\Gamma \Psi:=(\partial_x\Psi,\partial_y\Psi,0)^\tt$.\smallskip

\item For  $\Phib=(\Phib_x,\Phib_y,\Phib_z):\Gamma\to\mathbb{C}^3$ we set $\nabla_\Gamma\cdot \Phib:=\partial_x\Phib_x+\partial_y\Phib_y $.\smallskip

\item  $\mathsf{C}^m(\Omega) $, $m\leq\infty$, is the space  of functions with continuous derivatives up to order $m$  on  $\Omega $.\smallskip

\item  $\mathsf{C}^m(\overline\Omega) $, $m\leq\infty$, is the space  of 
all restrictions of functions in $\mathsf{C}^m(\mathbb{R}^n)$ to $\overline\Omega$.\smallskip

\item  $\mathsf{C}_0^m(\Omega)$ is the space  of  functions  $f\in \mathsf{C}^m(\Omega)$ having compact support in $\Omega$ (i.e., $\supp(f):=\overline{\{x:\ f(x)\not=0 \}}$ is a compact set contained in $\Omega$).\smallskip

\item  $\Dps(\Omega)$ is the space  of generalized scalar functions on $\Omega$ (i.e., linear continuous  functionals  acting on $\Cs(\Omega)$; see Appendix).\smallskip

\item $\Ls(\Omega)$  is the space of measurable functions $\Psi:\Omega\to\mathbb{C}$ satisfying $\intl_{\Omega}|\Psi(\r)|^2\d \r<\infty$.\smallskip

\item $\mathsf{H}^1(\Omega)$ is the space of measurable functions $\Psi:\Omega\to\mathbb{C}$ such that $\Psi\in \Ls(\Omega)$ and each component of $\nabla\Psi$ belongs to $\Ls(\Omega)$. Hereinafter,  differential operations   are understood in the generalized sense.\smallskip

\item $\mathbf{C}^m(\Omega)$ (resp. $\mathbf{C}^m(\overline\Omega) $, $\mathbf{C}_0^m(\Omega)$,   $\L(\Omega)$, $\H^1(\Omega)$) is the space of vector-functions $\Phib:\Omega\to\mathbb{C}^3$ with components being in  $\mathsf{C}^m(\Omega)$  (resp. $\mathsf{C}^m(\overline\Omega)$,  $\mathsf{C}_0^m(\Omega)$,   $\Ls(\Omega)$, $\mathsf{H}^1(\Omega)$).\smallskip

\item $\Dp(\Omega)$ is the space  of generalized vector functions on $\Omega$ (i.e., linear functionals  acting on $\C(\Omega)$).\smallskip 

\item $\Locs(\Omega)$ (resp. $\Loc(\Omega)$, $\mathsf{H}^1_{\rm loc}(\Omega)$, $\mathbf{H}^1_{\rm loc}(\Omega)$) is the space of (scalar or vector-valued) functions belonging 
to $\Ls(\widehat\Omega)$ (resp. $\L(\widehat\Omega)$, $\mathsf{H}^1 (\widehat\Omega)$, $\mathbf{H}^1 (\widehat\Omega)$) for each bounded subdomain $\widehat\Omega\subset \Omega$.
\smallskip

\item $\Id$ is the identity $(3\times 3)$-matrix .

\end{itemize}

\section{Maxwell's equations and the considered constitutive relations\label{sec1}}

We assume that the upper half-space $\Rp$ is occupied by 
vacuum, while the lower subspace $\Rm$ is occupied by a homogeneous metamaterial. We want to derive the interface conditions for this situation. Recall that $\Gamma$ is the interface between them.

As usual, we denote by $ \vec{\rm E}(\r,t)$, $ \vec{\rm B}(\r,t)$, $ \vec{\rm D}(\r,t)$ and $ \vec{\rm H}(\r,t)$ the electric field, the magnetic field,
the electric displacement field, and the magnetizing field. 
Assuming the absence of external charges and currents, we 
conclude that they obey  Maxwell's equations (in Gaussian units) in $\mathbb{R}^3\times\mathbb{R}$:
\begin{gather*}
\curl \vec{\rm E}+{1\over c}{\partial \vec{\rm B}\over\partial t}=0,\quad
\curl \vec{\rm H}-{1\over c}{\partial \vec{\rm D}\over\partial t}=0,\quad
\div \vec{\rm D}=0,\ 
\div \vec{\rm B}=0.
\end{gather*}

We are interested in monochromatic waves, i.e.,  waves of the form 
$$(\vec{\rm E},\vec{\rm B},\vec{\rm D},\vec{\rm H})=(\E,\B,\D,\H)\exp(-i\omega t),$$  
where $\omega>0$ is a frequency, and the vector functions $\E,\B,\D,\H$ depend only on the spatial variable $\r$. 
Plugging this ansatz into Maxwell's equations we arrive at
\begin{gather}\label{maxwell}
\curl \E-{i\omega\over c}\B=0,\quad
\curl \H+{i\omega\over c}\D=0,\quad
\div \D=0,\quad
\div  \B=0.
\end{gather}

Equations \ref{maxwell} must be supplemented by constitutive relations linking 
the ``induced'' fields $\D,\H$ with the ``physical'' fields $\E,\B$:  
$$
\D=\mathbf{E} + \mathbf{P}(\mathbf{E}),\qquad\qquad \mathbf{H}=
\mathbf{B}-\mathbf{M}(\mathbf{H}),
$$
where $\mathbf{P}$ and $\mathbf{M}$ are the polarization and the magnetization, respectively. Note that here we are concerned only with centro-symmetric metamaterials where electro-magnetic coupling effects are not encountered. Therefore, the polarization depends only on $\mathbf{E}$ and the magnetization depends only on $\mathbf{H}$. {The consideration of non-centro-symmetric materials, of course, can be integrated into the discussion but it does not affect our conclusions.}

We can assume that $\mathbf{M}\equiv 0$. This simply suggests that all the natural materials we are considering are non-magnetic in their intrinsic response. The magnetic effects we possibly observe always emerge from a nonlocal electric response. Under such assumption we easily get from Eqns.~\ref{maxwell} the following equation for $\mathbf{E}$:
\begin{gather}\label{wave}
\curl\curl\mathbf{E}={\omega^2 \over c^2}\left(\mathbf{E}+\mathbf{P}(\mathbf{E})\right).
\end{gather}

We assume that the upper half space is filled by vacuum. Hence, for $z>0$ one has $\mathbf{P}\equiv 0$, while for $z<0$ (i.e., in the metamaterial) we assume the following constitutive relations that resemble those already mentioned in Eqn.~\ref{const-relations_test}:
\begin{gather}\label{const-relations}
\mathbf{P}(\E)= (\eps -1)   \E+\curl \a\left(\curl\E\right)
+\suml_{j\in \{x,y,z\}}{\partial_j}\left(\b^j {\partial_j\E}\right)+
\curl\curl \g\left(\curl\curl
\E\right).
\end{gather}
Here $\eps,\a,\b^j,\g:\overline{\Rm}\to\mathbb{C}^{3\times 3}$ are matrix-functions with $C^\infty$-smooth and bounded entries. They depend on the frequency, i.e. the material properties are dispersive. The frequency dependency is suppressed here to simplify the notation. As we consider homogenous metamaterials, they do not depend on the spatial coordinates. Thus, the considered wave equation can be rewritten as
\begin{multline}\label{main-eq}
\curl\curl \E= 
{\omega^2\over c^2}\bigg(\tilde\eps \E+\curl \tilde\a\left(\curl\E\right)+
\suml_{j\in \{x,y,z\}}{\partial_j}\left(\tilde\b^j {\partial_j\E }\right)+
\curl\curl \tilde\g\left(\curl\curl
\E\right)\bigg),
\end{multline}
where 
\begin{gather*}
\tilde\eps=\begin{cases}\eps,&z<0,\\1,&z>0,\end{cases}\quad
\tilde\a=\begin{cases}\a,&z<0,\\0,&z>0,\end{cases}\quad
\tilde\b^j=\begin{cases}\b^j,&z<0,\\0,&z>0,\end{cases}\quad
\tilde\g=\begin{cases}\g,&z<0,\\0,&z>0.\end{cases}
\end{gather*}

\section{Generalised solutions\label{sec2}}

Up to now our considerations were rather formal. Now it is time to ask: in which sense has Eqn.\ref{main-eq} to be understood? Since the matrix-functions $\tilde\eps,\tilde\a,\tilde\b^j,\tilde\g$ are only piecewise continuous, we are not allowed to regard the differential expression on the right-hand-side of Eqn.~\ref{main-eq} in the classical sense. The natural idea is then to treat this equation in a suitable generalised sense. {From these considerations we can extract already specific requirements concerning the different functions, in particular concerning the spaces in which they are defined. This is helpful in the following sections as interface conditions can be found that require the mere analysis of the space in which these functions are living.}

Recall (see Appendix for more details) that the notation $\Dp(\R)$ stands for the space of generalised functions (distributions), i.e., linear functionals  acting continuously on $\C(\R)$.
For example, for $\E\in \Loc(\R)$, $\curl\E\in\Dp(\R)$ is defined by the action
$$(\curl\E)[\Phib]:=\intl_{\R}\E\cdot(\curl\Phib)\d \r,\ \Phib\in \C(\R),$$
which mimics partial integration. Also higher derivatives of $\E\in\Loc(\R)$ can be defined analogously, but the discontinuous function $\widetilde\alpha$ on the right-hand-side of Eqn.~\ref{main-eq} causes difficulties in mimicing partial integration (which would be no problem if $\widetilde\a$ was a $C^1$-function). We can solve this difficulty by requiring the additional regularity condition $\curl\E\in \Loc(\Rm)$. Under this assumption the product $\tilde\a(\curl \E)$ is a (regular) generalized function, which is defined by $\a (\curl \E)$ in $\Rm$ and $0$ in $\Rp$ and now  we are able to define its generalized $\curl$-derivative. Actually we require even $\curl\E\in \Loc(\R)$, in order to treat the \emph{left-hand-side} of \ref{main-inteq} appropriately when defining generalized solutions. Note that the property $\curl\E\in \Loc(\R)$ is not guaranteed by the requirement that $\curl \E$ is in $\Loc(\Rp)$ and in $\Loc(\Rm)$, but also (some kind of) continuity of the tangential component of $\E$ at the interface is needed.

Analogous considerations for the other terms in Eqn.~\ref{main-eq} lead to the following natural regularity assumptions:
\begin{gather}\label{regularity}
\E\in \Loc(\R),
\quad \curl\E\in \Loc(\R),
\quad
\E\in \mathbf{H}^1_{\rm loc}(\Rm),
\quad
\curl\curl \E\in \Loc(\Rm).
\end{gather}

Now, Eqn.~\ref{main-eq} is understood as an equality in $\Dp(\R)$.
By the definition of the generalized derivatives (see Appendix), mimicing partial integration in all occuring terms, it is equivalent to
\begin{multline}\label{main-inteq}
\forall\Phib\in \C(\R):\ \intl_{\R} (\curl\E)\cdot (\curl\Phib)\d \r =
{\omega^2 \over c^2} 
\intl_{\Rp} \E\cdot\Phib\d\r +
{\omega^2 \over c^2}\intl_{\Rm}\Big(\eps \E\cdot\Phib
\\
+\a\left(\curl\E\right)\cdot(\curl\Phib) 
-\suml_{j\in\{x,y,z\}}\b^j {\partial_j \E } \cdot {\partial_j \Phib }
+\g\left(\curl\curl\E\right)\cdot
(\curl\curl\Phib)\Big)\d \r .
\end{multline}
The vector-function $\E:\R\to\mathbb{C}$ is said to be a \textit{generalized} (or \textit{weak})
solution to Eqn.~\ref{main-eq} if it meets the regularity properties given in conditions \ref{regularity} and satisfies Eqn.~\ref{main-inteq}.
\smallskip

Note that one can introduce another definition of the generalized solution in which the requirement $\curl\E\in \Loc(\R)$ is omitted: we say that $\E:\R\to\mathbb{C}$ a \emph{very weak solution} to \ref{main-eq} if $\E\in \Loc(\R)$,
$\E\in \mathbf{H}^1_{\rm loc}(\Rm)$ {(implying also $\curl \E \in \L(\Rm)$)}, $\curl\curl \E\in \Loc(\Rm)$ and 
$
\forall\Phib\in \C(\R)$  $\intl_{\R} \E\cdot (\curl\curl\Phib)\d \r = \mathrm{RHS}_{\ref{main-inteq}}$,
where $\mathrm{RHS}_{\ref{main-inteq}}$ denotes the right-hand-side of equality \ref{main-inteq}. 
Evidently, if $\E$ is a weak solution to \ref{main-eq} then it is also a very weak solution (apparently the opposite is not true). In this paper we focus on weak solutions since for local constitutive relations (i.e., $\a=\gamma=\b^j=0$) they satisfy classical interface conditions
$[\E\times n],\ [\tilde\eps\E\cdot n],\
[(\curl\E)\times n],\ [(\curl\E)\cdot n]=0$ on $\Gamma$
(here $[\dots]$ stands for the jump of the enclosed quantity across $\Gamma$) -- this follows immediately from our results below. 
\smallskip

We also remark that analogous considerations can be performed for $\mathbf{L}^1$-type spaces.
Nevertheless, we prefer to deal with $\mathbf{L}^2$ functions and  spaces, since we expect that the $\mathbf{L}^2$ setting yields more benefits in subsequent research (in particular, since $\mathbf{L}^2$ is a Hilbert space).

Let $\E$ be a generalized solution to Eqn.~\ref{main-eq}. 
It is easy to show, just by taking $\Phib\in \C(\Rp)$ in Eqn.~\ref{main-inteq}, that 
\begin{gather}\label{ex1-}
\curl\curl \E\in  \Loc(\Rp) 
\end{gather}
and
\begin{gather}
\label{ex1}
\curl\curl \E= {\omega^2  \over c^2} \E \text{\,\, for almost all }x\in\Rp.
\end{gather}
``Almost all'' means that the set consisting of points $x\in\Rp$ at which the property expressed in Eqn.~\ref{ex1} fails has a Lebesque measure zero.

Similarly, taking $\Phib\in \C(\Rm)$, we conclude that
\begin{gather}\label{ex2-}
\curl \a\left(\curl\E\right)+ 
\suml_{j\in \{x,y,z\}}{\partial_j}\left(\b^j {\partial_j\E }\right)+
\curl\curl  \g\left(\curl\curl
\E\right)\in \Loc(\Rm) 
\end{gather}
and 
\begin{multline}\label{ex2}
\curl\curl \E={\omega^2  \over c^2} 
 \bigg( \eps \E+\curl \a\left(\curl\E\right) \\+
\suml_{j\in \{x,y,z\}}{\partial_j}\left(\b^j {\partial_j\E }\right)+
\curl\curl  \g\left(\curl\curl
\E\right)\bigg) 
\text{\,\, for almost all }x\in\Rm.
\end{multline}
Note that the regularity property given in expression \ref{ex2-} means that the sum of the generalized functions that appear in this expression are in $\Loc(\Rm)$, but it does not imply that the individual generalized functions belong to $\Loc(\Rm)$. 

Now that we have analyzed generalized solutions outside the interface $\Gamma$, we can derive the actual interface conditions in the next section.
 
\section{Derivation of the interface conditions\label{sec3}}

Let $\E$ be a generalised solution to Eqn.~\ref{main-eq}. We denote by $\E_+$ and $\E_-$ the restrictions of $\E$ to $\Rp$ and $\Rm$, respectively.
Also, we additionally assume in this section that $\E$ is smooth in each half-space, namely
\begin{gather}\label{regul}
\E_+\in C^4(\overline{\Rp}),\ \E_-\in C^4(\overline{\Rm}).
\end{gather}
These additional smoothness conditions are indeed satisfied if the coefficients  $\a,\b^s,\gamma $
are smooth, at least in the case where the differential Eqn.~\ref{main-eq} is elliptic (e.g., if $\gamma$ is positive definite); see \cite{F69}. The interface conditions we are going to derive in this section under the additional smoothness assumption expressed in \ref{regul} also remain valid when only our conditions \ref{regularity} are satisfied. But then they hold only in some generalized sense which needs the concept of traces. This will be outlined in Section~\ref{sec5}.

Below in volume integrals we will use the notation $\E$, while in integrals over $\Gamma$ we deal with $\E_+$ and $\E_-$. In the following we distinguish between the main interface conditions and two alternative interface conditions. These alternative interface conditions do not contain any further information and indeed can be derived from the main interface conditions. However, their documentation seems useful as these alternative interface conditions look simpler. This may make them more suitable for use in some specific situations. 

\subsection{Main interface conditions}

In the following we want to prove that if $\E$ is a generalized solution of Eqn.~\ref{main-eq} and satisfies the regularity assumptions \ref{regul}, then $\E$ satisfies the following interface conditions on $\Gamma$:
\begin{gather}
\label{cond1f}\tag{C$_1$}
(\E_+ -\E_-)\tn \,=\,0,
\\[1mm]\label{cond3f}\tag{C$_2$}
\left(\curl\E_+ - \curl\E_-\right)\tn\,
+\,{\omega^2\over c^2}\left(\a\curl\E_-
+\curl\gamma\curl\curl\E_- \right) \tn
-\,
{\omega^2\over c^2}(\Id-\n\n^{\tt})\b^z\partial_z\E_-
\, =\, 0,
\\[1mm]
\label{cond4f}\tag{C$_3$} 
\left(\gamma\curl\curl\E_-\right)\tn\,=\,0,
\\[1mm]
\label{cond5f}\tag{C$_4$}
\left(\b^z\partial_z\E_-\right)\cdot \n\, =\, 0.
\end{gather}

Conversely, if $\E$ satisfies \ref{regul}, solves \ref{ex1} in $\Rp$, solves \ref{ex2} in $\Rm$ and the conditions \ref{cond1f}-\ref{cond5f} are fulfilled, then $\E$ is a generalized solution to \ref{main-eq}. Please note, these interface conditions are one of the central results from our contributions. In the following we prove each of them.

\subsubsection{Proof of the first interface condition}
Since $\E\in\Loc(\R)$ and $\curl\E\in\Loc(\R)$ we have
\begin{gather}\label{cond1.1}
\forall\Phib\in\C(\R):\ \intl_{\R}  \E\cdot (\curl\Phib ) \d \r=
\intl_{\R}(\curl\E )\cdot \Phib \d \r.
\end{gather}
On the other hand, by integrating by parts in each half-space we obtain
\begin{multline}\label{cond1.2}
\intl_{\R}(\curl\E )\cdot \Phib \d \r=
\intl_{\Rm}(\curl\E) \cdot \Phib \d \r+
\intl_{\Rp}(\curl\E)\cdot \Phib \d \r \\=
\intl_{\Rm}\E \cdot (\curl\Phib) \d \r+
\intl_{\Rp}\E \cdot (\curl\Phib) \d \r+
\intl_{\Gamma}\left(\E_+\tn - \E_-\tn\right)\cdot \Phib \d \s \\=
\intl_{\R}\E  \cdot (\curl\Phib) \d \r+
\intl_{\Gamma}\left(\E_+\tn - \E_-\tn\right)\cdot \Phib \d \s,
\end{multline}
where $\d \s=\d x \d y$ is the area of the surface element on $\Gamma$. Since $\Phib$ is arbitrary, we obtain from Eqns.~\ref{cond1.1} and \ref{cond1.2} the first interface conditions \ref{cond1f}.

\subsubsection{Proof of the second interface condition}
To proof the second interface condition we  decompose each integral in Eqn.~\ref{main-inteq} in a sum of integrals over $\Rm$ and $\Rp$ and integrate by parts in such a way that all the derivatives shift from $\Phib$ to $\E$. 
Then, moving all volume integrals to the left-hand-side and all integrals over $\Gamma$ to the right-hand-side, we get
\begin{multline}\label{cond3.1}
\intl_{\R} (\curl\curl\E)\cdot \Phib\d \r -
{\omega^2 \over c^2} 
\intl_{\Rp} \E\cdot\Phib\d\r -
{\omega^2 \over c^2}\intl_{\Rm}\Big(\eps \E
\\
+\curl\a\curl\E 
+\suml_{j\in\{x,y,z\}}\partial_j\left(\b^j {\partial_j \E }\right)  
+\curl\curl\g\left(\curl\curl\E_-\right)\Big)\cdot
\Phib\d \r  \\=
\intl_\Gamma \left(\left(\curl\E_+ - \curl\E_-  
+{\omega^2\over c^2}\left(\a\curl\E_-
+\curl\gamma\curl\curl\E_-\right)\right)\tn\right)\cdot\Phib\d\s\\+
{\omega^2\over c^2}\intl_\Gamma 
\left(\left(\gamma\curl\curl\E_-\right)\tn\right)\cdot(\curl\Phib)\d\s-
{\omega^2\over c^2}\intl_\Gamma 
\left(\b^z\partial_z\E_-\right)\cdot \Phib\d\s.
\end{multline}
Due to Eqns.~\ref{ex1} and \ref{ex2}, the left-hand-side of Eqn.~\ref{cond3.1} vanishes and thus Eqn.~\ref{cond3.1} can be rewritten as follows:
\begin{multline}\label{cond3.2}
\intl_\Gamma 
\left(
\left(\curl\E_+ - \curl\E_-  
+{\omega^2\over c^2}\left(\a\curl\E_-
+\curl\gamma\curl\curl\E_- \right)\right)\tn-
{\omega^2\over c^2}(\Id-\n\n^{\tt})\b^z\partial_z\E_- 
\right)
\cdot\Phib\d\s\\+
{\omega^2\over c^2}\intl_\Gamma 
\left(\left(\gamma\curl\curl\E_-\right)\tn\right)\cdot(\curl\Phib)\d\s-
{\omega^2\over c^2}\intl_\Gamma 
\left(\n\n^{\tt}\b^z\partial_z\E_-\right)\cdot \Phib\d\s\, =\, 0.
\end{multline}

Now, we choose the function $\Phib$ of the form
\begin{gather}\label{Phi1}
\Phib(\r)=(\Phi_1(x,y)\eta(z),\,\Phi_2(x,y)\eta(z),\,0)^\tt,
\end{gather}
where $\Phi_1,\Phi_2\in \Cs(\mathbb{R}^2)$, $\eta\in \Cs(\mathbb{R})$, moreover $\eta|_{\{|z|<\delta\}}=1$  for some $\delta>0$.
One has:
$\curl\Phib=(-\Phi_2\eta',\,\Phi_1\eta',\,(\partial_x\Phi_2-\partial_y\Phi_1)\eta)^\tt$,
whence $$\curl\Phib|_{\Gamma}=(0,\,0,\ \partial_x\Phi_2-\partial_y\Phi_1)^\tt,$$
and, hence, the second integral in Eqn.~\ref{cond3.2} vanishes. Since the $z$-component of $\Phib$ is equal to zero, the third integral in Eqn.~\ref{cond3.2} also vanishes.
Thus  
\begin{multline}\label{cond3.3}
\intl_\Gamma 
\left(
\left(\curl\E_+ - \curl\E_-  
+{\omega^2\over c^2}\left(\a\curl\E_-
+\curl\gamma\curl\curl\E_- \right)\right)\tn\right.\\-\left.
{\omega^2\over c^2}(\Id-\n\n^{\tt})\b^z\partial_z\E_-
\right)
\cdot(\Phi_1,\Phi_2,0)^{\tt}\d\s\, =\, 0.
\end{multline}
Finally, since the functions $\Phi_1$ and $\Phi_2$ are arbitrary, 
we conclude from Eqn.~\ref{cond3.3} the interface condition \ref{cond3f}.
 
\subsubsection{Proof of the third interface condition}

To proof the third interface condition we take the function $\Phib$ of the form
\begin{gather}\label{Phi2}
\Phib(\r)=(\Phi_2(x,y)z\eta(z),\,-\Phi_1(x,y)z\eta(z),\,0)^\tt.
\end{gather}
As before, $\Phi_1,\Phi_2\in \Cs(\mathbb{R}^2)$, $\eta\in  \Cs(\mathbb{R})$,  and $\eta|_{\{|z|<\delta\}}=1$  for some $\delta>0$.
We get
$\curl\Phib=((\eta+z\eta')\Phi_1,\, (\eta+z\eta')\Phi_2,\,(-\partial_x\Phi_1-\partial_y\Phi_2)z\eta)^\tt$ and hence
\begin{gather}\label{cond4.1}
\Phib|_{\Gamma}=0,\quad \curl\Phib|_{\Gamma}=(\Phi_1,\,\Phi_2,\ 0)^\tt.
\end{gather}
We plug this function $\Phib$ into Eqn.~\ref{cond3.2}. Due to \ref{cond4.1},
all integrals in Eqn.~\ref{cond3.2} vanish except the second one. Thus one gets:
\begin{gather}\label{cond4.2}
\intl_\Gamma 
\Big(\left(\gamma\curl\curl\E_-\right)\tn\Big)\cdot (\Phi_1,\Phi_2,0)^\tt\d\s\, =\, 0.
\end{gather}
Since $\Phi_1$ and $\Phi_2$ are arbitrary, Eqn.~\ref{cond4.2} implies the third interface condition \ref{cond4f}.

\subsubsection{Proof of the fourth interface condition}

Finally, due to \ref{cond3f}, \ref{cond4f}, the first two integrals in Eqn.~\ref{cond3.2} vanish for any arbitrary function $\Phib$, whence one gets
\begin{gather}\label{cond5.1} 
\intl_\Gamma 
\left(\n\n^{\tt}\b^z\partial_z\E_- \right)\cdot \Phib\d\s\, =\, 0,
\end{gather}
and, consequently, we arrive at the last condition \ref{cond5f}.\medskip

\subsubsection{Further remarks}
Conversely, let now $\E$ satisfy requirement~\ref{regul},  solve Eqn.~\ref{ex1} in $\Rp$, solve Eqn.~\ref{ex2} in $\Rm$, and let \ref{cond1f}--\ref{cond5f} hold. We have to show that $\E$ is a generalized solution to Eqn.~\ref{main-eq}. 

Indeed, it follows from the assumption~\ref{regul} that $\E\in\Loc(\R)$, $\E\in\mathbf{H}^1_{\rm loc}(\Rm)$ and $\curl\curl\E\in\Loc(\Rm)$. Moreover, due to \ref{cond1f}, $\curl\E\in\Loc(\R)$. 
Thus, conditions \ref{regularity} are satisfied. 

Integrating by parts we get for an arbitrary $\Phib\in\C(\R)$:
\begin{multline}
\intl_{\R} (\curl\E)\cdot (\curl\Phib)\d \r -
{\omega^2 \over c^2} 
\intl_{\Rp} \E\cdot\Phib\d\r -
{\omega^2 \over c^2}\intl_{\Rm}\Big(\eps \E\cdot\Phib
+\a\left(\curl\E\right)\cdot(\curl\Phib)\\
-\suml_{j\in\{x,y,z\}}\b^j {\partial_j \E } \cdot {\partial_j \Phib }+\g\left(\curl\curl\E\right)\cdot
(\curl\curl\Phib)\Big)\d \r = \mathrm{LHS}_{\ref{cond3.1}}- \mathrm{RHS}_{\ref{cond3.1}}.
\end{multline}
where $\mathrm{LHS}_{\ref{cond3.1}}$ (resp. $\mathrm{RHS}_{\ref{cond3.1}}$) is the expression standing in the left-hand-side (resp. right-hand-side) of  Eqn.~\ref{cond3.1}.
Due to \ref{ex1}-\ref{ex2}, $\mathrm{LHS}_{\ref{cond3.1}}=0$ and, due to conditions \ref{cond3f}-\ref{cond5f}, $\mathrm{RHS}_{\ref{cond3.1}}=0$.  Therefore, 
Eqn.~\ref{main-inteq} holds true and, consequently, $\E$ is a generalized solution to Eqn.~\ref{main-eq}.

With that we have been offering proofs for the main interface conditions. In the following, we formulate two alternative interface conditions that can be used as well, but they are not fundamental since they follow from the definition of the generalized solution and the main interface conditions.  

\subsection{Alternative interface conditions}

Let $\E$ be a generalized solution to Eqn.~\ref{main-eq} satisfying the regularity assumptions 
\ref{regul}. Then $\E$ meets the interface conditions  \ref{cond1f}-\ref{cond5f}. The fulfilment of \ref{cond1f}-\ref{cond5f} is also a sufficient condition for being a generalized solution if \ref{ex1} and \ref{ex2} hold -- see the statement after conditions \ref{cond1f}-\ref{cond5f}. 
In this subsection, we derive two alternative interface conditions on $\Gamma$. They appear slightly simpler and are, therefore, of practical use in further research.

\subsubsection{First alternative interface condition}

Let $\Psi\in\Cs(\R)$ be an arbitrary function.
Since $\curl\E\in \Loc(\R)$,
$\div(\curl \E)=0$ and $\curl(\nabla\Psi)=0$ one has
\begin{multline*}
0=-\intl_{\R}(\div(\curl \E))\Psi\d\r=-\intl_{\Rp}(\div(\curl \E))\Psi\d\r-\intl_{\Rm}(\div(\curl \E))\Psi\d\r\\=
\intl_\Gamma ( \left(\curl \E_+ - \curl \E_- \right)\cdot \n)\Psi\d\s+ \intl_{\Rp}(\curl \E)\cdot(\nabla \Psi)\d\r+\intl_{\Rm}(\curl \E)\cdot(\nabla \Psi)\d\r\\=
\intl_\Gamma ( \left(\curl \E_+ - \curl \E_- \right)\cdot \n)\Psi\d\s+
\intl_\Gamma ( \left(\E_+ - \E_- \right)\tn)\cdot\nabla\Psi\d\s+
\intl_{\R}\E\cdot (\curl(\nabla \Psi))\d\r\\\overset{\text{\ref{cond1f}}}{=}
\intl_\Gamma ( \left(\curl \E_+ - \curl \E_- \right)\cdot \n)\Psi\d\s,
\end{multline*}
whence we get the following interface condition on  $\Gamma$:
\begin{gather}\label{cond2f}\tag{C$_5$}
\left(\curl \E_+ - \curl \E_- \right)\cdot \n=0.
\end{gather}

\subsubsection{Second alternative interface condition}

For materials governed by local constitutive relations  (i.e., $\a=\gamma=\b^j=0$) one has 
also the interface condition 
\begin{gather}\label{cond-add1}
\left( \E_+ - \eps\E_- \right)\cdot \n=0,
\end{gather}
which follows from the fact that $\tilde\eps\E\in \Loc(\R)$ and, due to Eqn.~\ref{main-eq}, $\div (\tilde\eps\E)=0$. 

Let us  now derive an analogue of the interface condition in Eqn.~\ref{cond-add1} for our nonlocal model.
For this purpose, we plug into Eqn.~\ref{main-inteq}  the function $\Phib$ of the form
$$\Phib=\nabla\Psi ,$$
where $\Psi$ is a smooth compactly supported scalar function. Under this choice $\curl\Phib=0$ and
Eqn.~\ref{main-inteq} becomes
\begin{gather}\label{cond-add2}
\intl_{\Rp} \E\cdot\nabla\Psi \d\r +
\intl_{\Rm}\left(\eps \E\cdot\nabla\Psi 
-\suml_{j\in\{x,y,z\}}\b^j {\partial_j \E} \cdot {\partial_j \nabla\Psi }\right)
 \d \r =0.
\end{gather}
Let us additionally assume that $\mathrm{supp}\Psi \subset\Rm$;
integrating by part in Eqn.~\ref{cond-add2} we get:
\begin{gather*}
\forall \Psi \in \Cs(\Rm):\ 
\intl_{\Rm}\left(\div\left(\eps \E
+\suml_{j\in\{x,y,z\}}\partial_j\left(\b^j \partial_j \E \right) \right)\right)
\Psi \d \r =0,
\end{gather*}
whence, since $\Cs(\Rm)$ is dense in $\Ls(\Rm)$,
\begin{gather}\label{cond-add4}
\div\left(\eps \E 
+\suml_{j\in\{x,y,z\}}\partial_j\left(\b^j \partial_j \E \right) \right)=0\text{\,\, in }\Rm.
\end{gather}
Similarly, by using $\Psi \in \Cs(\Rp)$, we obtain: 
\begin{gather}\label{cond-add5}
\div \E=0\text{\,\, in }\Rp.
\end{gather} 

Now, we take an arbitrary $\Psi\in \Cs(\R)$ in Eqn.~\ref{cond-add2} and integrate by parts in each half-space $\R_\pm$. We get:
\begin{multline}\label{cond-add6}
\intl_{\Rp} (\div\E)\,  \Psi\d\r 
+
\intl_{\Rm}\div\left(
\eps \E +
\suml_{j\in\{x,y,z\}}\partial_j\left(\b^j {\partial_j \E }\right)\right)\,\Psi  \d \r\\
+\intl_{\Gamma}\left(\left( \E_+ - \eps\E_- 
-\suml_{j\in\{x,y,z\}}\partial_j\left(\b^j {\partial_j \E_- }\right) \right)\cdot \n\right) \Psi\d \s+
\intl_{\Gamma}\b^z\partial_z\E_-\cdot\nabla\Psi\d \s
=0.
\end{multline}
In this equality the integrals over $\R_\pm$ vanish due to Eqn.~\ref{cond-add4} and Eqn.~\ref{cond-add5}. Moreover, in view of interface condition \ref{cond5f}, the last integral in Eqn.~\ref{cond-add6} 
can be rewritten as follows:
\begin{multline}\label{cond-add7}
\intl_{\Gamma}\b^z\partial_z\E_-\cdot\nabla\Psi\d \s=\intl_{\Gamma}(\Id-\n\n^{\tt})\b^z\partial_z\E_-\cdot\nabla_\Gamma\Psi\d \s=
-\intl_{\Gamma}\nabla_\Gamma\cdot\left(\b^z\partial_z\E_-\right)\,\Psi\d \s.
\end{multline}
Since $\Psi \in \Cs(\R)$ is an arbitrary function  we conclude
from Eqns.~\ref{cond-add6}-\ref{cond-add7} the following interface condition on  $\Gamma$:
\begin{gather}\label{cond-add}\tag{C$_6$}
\left( \E_+ - \eps\E_- 
-\suml_{j\in\{x,y,z\}}\partial_j\left(\b^j {\partial_j \E_- }\right) \right)\cdot \n-{ \nabla_\Gamma\cdot\left(\b^z\partial_z\E_z\right)}=0.
\end{gather}

That closes this section. In Section~\ref{sec5} we will clarify how to treat the interface conditions in the general case, i.e., when the generalized solution does not have the additional regularity expressed in condition \ref{regul}. 

We will use these interface conditions to derive Fresnel equations for the transmission and reflection of a plane wave from a slab in the next section.

\section{Fresnel formulae\label{sec4}}
 
In this section, we apply the interface conditions obtained above to the problem of light propagation through a slab of metamaterial. 

The geometry of the pertinent problem hence is defined as $$\Omega=\left\{\r\in\R:\ -\delta<z<0 \right\},\quad
\Omega_-=\left\{\r\in\R:\ z<-\delta \right\},\quad
\Omega_+=\left\{\r\in\R:\ z>0 \right\},$$
where $\delta>0$.
We assume that $\Omega$ is filled with a metamaterial, which is governed 
by the constitutive relations expressed in Eqn.~\ref{const-relations}. Moreover, additionally we assume that the coordinate system of the laboratory corresponds to the coordinate system of the principle axis of the metamaterial. The materials considered here are centro-symmetric. Therefore, the material tensors are considered to be diagonal. In essence, this means that the unit cells are aligned to the slab we consider. In this coordinate system the material properties read as 
\begin{gather}\label{coeff}
\begin{array}{lll}
\eps =
\left(
\begin{matrix}
\eps_x&0&0\\0&\eps_y&0\\0&0&\eps_z
\end{matrix}\right),&
\a =
\left(
\begin{matrix}
\a_x&0&0\\0&\a_y&0\\0&0&\a_z
\end{matrix}\right),&
\gamma =
\left(
\begin{matrix}
\gamma_x&0&0\\0&\gamma_y&0\\0&0&\gamma_z
\end{matrix}\right),
\\[7mm]
\beta^x =\left(\begin{matrix}
\beta_x&0&0\\0&0&0\\0&0&0
\end{matrix}\right),& \beta^y =\left(\begin{matrix}
0&0&0\\0&\beta_y&0\\0&0&0
\end{matrix}\right),& \beta^z =\left(\begin{matrix}
0&0&0\\0&0&0\\0&0&\beta_z
\end{matrix}\right).
\end{array}
\end{gather} 
Note that this is not an essential condition but it helps us to keep the expressions sufficiently simple. The remaining space (that is $\Omega_-\cup\Omega_+$) is occupied by vacuum.

We also denote 
$$\Gamma_-=\{\r\in\partial\Omega:\ z=-\delta\},\quad \Gamma_+=\{\r\in\partial\Omega:\ z=0\}.$$

Now, assume that we have an incident plane wave impinging on the the slab from $\Omega_+$.
A part of this wave will be reflected, the other part will be transmitted through the slab to 
$\Omega_-$. Our goal is to  find the amplitudes of these reflected and transmitted waves, 
in other words, we want to derive Fresnel-type formulae. Even though discussed here for a plane wave, an arbitrary illumination can always be written as a superposition of plane waves. Therefore, the plane wave assumption is by no means a limitation.

We notice that, due to the special form of the material coefficients expressed in Eq.~\ref{coeff},  
each solution $\E$ of Eqn.~\ref{main-eq} can be represented in the form 

$$\E=\E^{\TE}+\E^{\TM},$$ 
where $\E^{\TE}=(E^{\TE}_x,0,0)^\tt$ (transverse-electric-polarized wave) and $\E^{\TM}=(0,E^{\TM}_y,E^{\TM}_z)^\tt$ (transverse-magnetic-polarized wave), each of which satisfies Eqn.~\ref{main-eq}. In what follows, we treat $\TE$- and $\TM$-polarized incident waves separately.

In this section $\r$ is treated as vector-column, i.e. $\r=(x,y,z)^\tt$.

\subsection{$\TE$-polarization}

Assume that 
$$\E^I= \E_0^I\exp(\i\k^I\cdot \r)$$ 
is the incident $\TE$-polarized plane wave. Here $\k^I=(k^I_x,k^I_y,k^I_z)^\tt$ is the wave vector  and $\E_0^I=(E_x^I,0,0)^\tt$ is the amplitude vector of the incident plane wave. The wave vector in vacuum obeys the ordinary dispersion relation, i.e. $k^2=\frac{\omega^2}{c^2}$.
Moreover, in view of \ref{cond-add5} one has $k_x^I=0$.

Due to symmetry arguments it is reasonable to search 
the reflected and the transmitted fields in the same form as the incident field. Namely, the reflected field is searched in the form $$\E^R= \E_0^R\exp(\i\k^R\cdot \r),\text{ where } \k^R=(0,k^R_y,k^R_z)^\tt,\ \E_0^R=(E_x^R,0,0)^\tt.$$ 
The total field in $\Omega_+$ is
$\E^I+\E^R.$
In $\Omega_-$ we have the transmitted field
$$\E^T= \E_0^T\exp(\i\k^T\cdot \r),\text{ where }  \k^T=(0,k^T_y,k^T_z)^\tt,\ \E_0^T=(E_x^T,0,0)^\tt.$$ 
Finally, in the slab $\Omega$ the total field has the form
$$\E^{\text{slab}}= \suml_{j=1}^N  \E_0^{j}\exp(\i\k^{j}\cdot \r),\quad \k^{j}=(0,k^{j}_y,k^{j}_z)^\tt,\ \E_0^{j}=(E_x^{j},0,0)^\tt,$$ where $N$ is the number of linearly independent eigenmodes existing in $\Omega$. The larger number of plane waves is reminiscent to the fact that for each value of $k_x$ and $k_y$ multiple solutions for $k_z$ exist at each frequency. This will be discussed below. Also, the field inside the slab is always written as a superposition of forward and backward propagating modes in the principle propagation direction. Therefore, this quite general ansatz is chosen.

Plugging the plane wave ansatzes, for example, into \ref{cond1f} on $\Gamma_+$ and $\Gamma_-$ we get the equation
\begin{gather*}
E^I\e^{\i (k^I_xx+ k^I_yy)}+E^R\e^{\i (k^R_xx+ k^R_yy)}-\suml_{j=1}^N E_x^j \e^{\i (k^j_xx+ k^j_yy)}=0,\quad
\suml_{j=1}^N E_x^j\e^{-\i k^j_z\delta} \e^{\i (k^j_xx+ k^j_yy)}-E^T\e^{-\i k^T_z\delta}\e^{\i (k^T_xx+ k^T_yy)}=0
\end{gather*}
which hold for all $(x,y)\in\mathbb{R}^2$.  As the system is translational invariant along the interface, we require all plane waves involved in the process to share the same wave vector components tangential to the interface.
Hence the vectors $\k^I$, $\k^R$, $\k^T$ and $\k^{j}$ ($j=1,\dots,N$) have the same $y$-components. 
Hereafter for the $y$-component of all wave vectors we use the notation $k_y$.

It is easy to see that $N=4$ provided $\gamma_x\not=0$. Indeed, plugging $\E=(E_x,0,0)^\tt\exp(\i\k\cdot \r)$ (with $k_x=0$) into Eqn.~\ref{wave} supplemented with the constitutive relations expressed in Eqn.~\ref{const-relations}, we arrive at the following dispersion relation linking $\omega$ and $k$ for the metamaterial:
\begin{gather}\label{DR-TE}
k_y^2+k_z^2=\omega^2 c^{-2} \left(\eps_x+(\a_z k_y^2+\a_y k_z^2)+\gamma_x(k_y^2+k_z^2)^2\right).
\end{gather}
This is a 4$^\mathrm{th}$-order polynomial equation with respect to $k_z$, thus generically we get 
four eigenmodes~\footnote{We omit the special case, when Eqn.~\ref{DR-TE} has multiple roots. This case requires a separate treatment.}. They come in pairs and two of these eigenmodes are forward and two of these eigenmodes are backward propagating.

There are six unknowns $E_x^R$, $E_x^T$, $E_x^j$ ($j=1,\dots,4$).
On each interface $\Gamma_+$ and $\Gamma_-$ we can initially impose four conditions \ref{cond1f}-\ref{cond5f}, but   \ref{cond5f} simply reads   $0=0$.
As a result, we have three non-trivial equations on each interface. This leaves us with a total number of six linearly independent equations which is just enough to solve uniquely for all involved amplitudes.

Plugging our plane wave ansatzes into these equations we arrive at  the following linear algebraic system for ${E}=(E_x^R, E_x^1, E_x^2, E_x^3, E_x^4, E_x^T)^\tt$: 
$$\mathcal{A}{E}=\mathcal{F},$$
where 
\begin{gather*} 
\mathcal{A}=
\left( 
\begin{array}{cccccc}
1& 
-1&
-1&
-1&
-1&
0
\\[5mm]
\ds k_z^R  &
 -A_1 &
 -A_2 &
 -A_3 &
 -A_4 &
0
\\[5mm]
0& 
{\gamma_x}(\k^1)^2 &
{\gamma_x}(\k^2)^2 &
{\gamma_x}(\k^3)^2 &
{\gamma_x}(\k^4)^2 &
0\\[5mm]
0& 
-\ds\e^{-\i k_z^1 \delta}&
-\ds\e^{-\i k_z^2 \delta}&
-\ds\e^{-\i k_z^3 \delta}&
-\ds\e^{-\i k_z^4 \delta}&
 \ds\e^{-\i k_z^T \delta}
\\[5mm]
\ds 0 &
-A_1\ds\e^{-\i k_z^1 \delta}&
-A_2\ds\e^{-\i k_z^2 \delta}&
-A_3\ds\e^{-\i k_z^3 \delta}&
-A_4\ds\e^{-\i k_z^4 \delta}&
k_z^T\ds\e^{-\i k_z^T \delta}
\\[5mm]
0& 
{\gamma_x}(\k^1)^2\ds\e^{-\i k_z^1 \delta}&
{\gamma_x}(\k^2)^2\ds\e^{-\i k_z^2 \delta}&
{\gamma_x}(\k^3)^2\ds\e^{-\i k_z^3 \delta}&
{\gamma_x}(\k^4)^2\ds\e^{-\i k_z^4 \delta}&
0\\[5mm]
\end{array}
\right),\quad
\mathcal{F}=
-E_x^I
\left(
\begin{matrix}
1\\[5mm]
\ds k_z^I 
\\[5mm]0\\[5mm]0
\\[5mm]0\\[5mm]0
\end{matrix}\right).
\end{gather*}
Here $(\k^j)^2:=(k_y)^2+(k^j_z)^2$, and 
$\ds A_j:=k_z^j\left(1-{\omega^2\over c^2}\left(\a_y+\g_x(\k^j)^2\right)\right)$.
Thus the required amplitudes $E_x^R$ and $E_x^T$ are determined by
$$ 
E^R=\left[\left(\mathcal{A}^{-1}\mathcal{F}\right)\right]_1,\quad
E^T=\left[\left(\mathcal{A}^{-1}\mathcal{F}\right)\right]_6,$$
where $[\cdot]_k$ denotes the $k$-th component of a vector. {It is not hard to show that generically  the matrix $\mathcal{A}$ is invertible; ``generically'' means that 
the set consisting of $\eps,\alpha,\gamma,\beta^j$ for which $\mathrm{det}(\mathcal{A})=0$ has Lebesgue measure zero in the space of all admissible parameters}.

Note, that the equation coming from conditions \ref{cond2f} coincides with the equation coming from condition \ref{cond1f} multiplied by $k_y$, and the equation coming from condition \ref{cond-add} reads $0=0$. 

\subsection{$\TM$-polarization}

Again, in $\Omega_+$ the total field is of the form
$\E^I+\E^R$,
where $\E^I=\E_0^I\exp(\i\k^I\cdot \r)$ is the incident 
field, $\E^R=\E_0^R\exp(\i\k^R\cdot \r)$ is the reflected field, but now
the fields are $\TM$-polarized, i.e.,
$\E_0^I=(0,E_y^I,E_z^I)^\tt$, $\E_0^R=(0,E_y^R,E_z^R)^\tt$.
In $\Omega_-$ we get the transmitted field
$\E^T= \E_0^T\exp(\i\k^T\cdot \r)$, where $\E_0^T=(0,E_y^T,E_z^T)^\tt.$ 
Finally, in the slab $\Omega$ the total field has the form
$\E^{\text{slab}}=\suml_{j=1}^N  \E_0^{j}\exp(\i\k^{j}\cdot \r)$, where $ \E_0^{j}=(0,E_y^j,E_z^{j})^\tt.$

Again the vectors $\k^I$, $\k^R$, $\k^T$ and $\k^{j}$ ($j=1,\dots,N$) have the same $x$- and $y$-components. Similarly to the $\TE$-polarization, we denote by $k_y$ the $y$-component of 
the wave vectors, the $x$-components are chosen to be equal to zero.

For $\TM$-polarization the dispersion relation reads
\begin{gather}
\label{dr-tm}
\mathrm{det}\left(\begin{matrix}-\omega^2 c^{-2}\eps_y+
k_z^2 Q+ \omega^2  c^{-2}\b_y k_y^2&
-k_y k_z Q
\\[3mm]-k_y k_zQ &-\omega^2 c^{-2}\eps_z+
k_y^2Q+ \omega^2 c^{-2}\b_z k_z^2\end{matrix}\right)=0,
\end{gather}
where $Q:=1-\omega^2 c^{-2}(\a_x+\gamma_z k_y^2+\gamma_y k_z^2)$.
It is easy to see that this is a 6$^\mathrm{th}$-order polynomial equation with respect to $k_z$
provided $\gamma_y\not=0$, $\beta_z\not=0$. Thus $N=6$. There are three forward and three backward propagating modes.

In $\Omega_+\cup\Omega_-$ one has $\mathrm{div}\,\mathbf{E}=0
$ and consequently
\begin{gather}\label{d1}
E_yk_y   +E_zk_z = 0.
\end{gather}
In $\Omega$ one has $\mathrm{div}\left(\eps  \mathbf{E}+\ds\suml_{j\in\{x,y,z\}}\partial_j(\beta^j\partial_j\mathbf{E}
)\right)=0$  and consequently
\begin{gather}\label{d2}
E_yk_y(\eps_y-\beta_y k_y^2)+E_zk_z(\eps_z-\beta_z k_z^2) =0.
\end{gather}
Due to Eqns.~\ref{d1}-\ref{d2}, it is enough to determine only the $z$-components of the reflected and transmitted fields (except in the special case $(\eps_y - \beta_y k_y^2)k_y=0$, which we do not treat here).

We have eight unknowns $E_z^R$, $E_z^T$, $E_z^j$ ($j=1,\dots,6$).
On each interface $\Gamma_+$ and $\Gamma_-$ we have  four conditions \ref{cond1f}-\ref{cond5f}, thus, both interfaces produce totally eight equations.
Plugging the plane wave ansatzes  into these equations
we arrive at the following system for $ {E}=(E_z^R, E_z^1, E_z^2, E_z^3, E_z^4, E_z^5, E_z^6,  E_z^T)^\tt$: $\mathcal{A} {E}=\mathcal{F},$
where \footnotesize
\begin{gather*} 
\mathcal{A}=\overset{\hspace{30mm}j=1\hspace{36mm} j=2\hspace{2mm} j=3\hspace{2mm} j=4\hspace{2mm} j=5\hspace{2mm}  j=6}{\left(
\begin{array}{c|c|c|c|c|c|c|c}
k_z^R & 
-B_1k_z^1 &
\dots&
\dots&
\dots&
\dots&
\dots&
0
\\[5mm]
 (\k^R)^2 &
\left(-1+{\omega^2 \over c^2}\a_y+{\omega^2 \over c^2}\left(k_y^2\gamma_z+(k_z^1)^2\gamma_y\right)\right)\left(k_y^2+(k_z^1)^2
B_1
\right)   &
\dots&
\dots&
\dots&
\dots&
\dots&
0
\\[5mm]
0& 
{\gamma_y}k_z^1\left(  k_y^2+(k_z^1)^2B_1\right) &
\dots&
\dots&
\dots&
\dots&
\dots&
0\\[5mm]
0& {\beta_z}k_z^1 &
\dots&
\dots&
\dots&
\dots&
\dots&
0
\\[5mm]
0& 
-\ds B_1 k_z^1 \e^{-\i k_z^1 \delta} &
\dots&
\dots&
\dots&
\dots&
\dots&
 k_z^T   \e^{-\i k_z^T \delta} 
\\[5mm]
0 & \left(-1+{\omega^2 \over c^2}\a_y+{\omega^2 \over c^2}\left(k_y^2\gamma_z+(k_z^1)^2\gamma_y\right)\right)\left(k_y^2+(k_z^1)^2
B_1
\right)  \e^{-\i k_z^1 \delta} 
 &
\dots&
\dots&
\dots&
\dots&
\dots&(\k^T)^2  \e^{-\i k_z^T \delta} 
\\[5mm]
0& 
{\gamma_y}k_z^1\left(  k_y^2+(k_z^1)^2B_1\right)  \e^{-\i k_z^1 \delta} &
\dots&
\dots&
\dots&
\dots&
\dots&
0\\[5mm]
0& {\beta_z}k_z^1  \e^{-\i k_z^1 \delta} &
\dots&
\dots&
\dots&
\dots&
\dots&
0
\end{array}
\right)},\\[3mm]
\mathcal{F}=-E_z^I\left(
 k_z^I,\
 (\k^I)^2,\
0,\
0,\
0,\
0,\
0,\
0
\right)^{\tt},
\end{gather*}\normalsize
and $B_j:=\ds{\eps_z -(k_z^j)^2\beta_z\over\eps_y - (k_y)^2 \beta_y}$,
$(\k^*)^2:=(\k_y)^2+(\k_z^*)^2$. In the matrix $\mathcal{A}$, the entries of the columns indicated  by $j=2,3,4,5,6$ have the entries as the column with $j=1$, but with $k_z^j$ instead of $k_z^1$. 

{Again one can show that generically  the matrix $\mathcal{A}$ is invertible.} The required amplitudes $E_z^R$ and $E_z^T$ are determined by
$$
E^R=\left[\left(\mathcal{A}^{-1}\mathcal{F}\right)\right]_1,\quad
E^T=\left[\left(\mathcal{A}^{-1}\mathcal{F}\right)\right]_8.$$
\smallskip

We note that for $\TM$-polarization the alternative condition \ref{cond2f} reads  $0=0$, while the equations
generated by \ref{cond-add} on $\Gamma_+$ and $\Gamma_0$ can be deduced from  
the equations generated by the main interface conditions and the dispersion relations \ref{dr-tm}.
\medskip

The derivation of the Fresnel coefficients is the second major achievement of this work. They are explicitly documented in this work and can be used right away for further research.

\section{Traces\label{sec5}}

Let $\E$ be a generalized solution to Eqn.~\ref{main-eq}, i.e., \ref{regularity} and \ref{main-inteq} hold.
Since $\E$ belongs to $\Loc(\R)$, it is defined up to a set of Lebesgue measure zero.
The interface $\Gamma$ has Lebesgue measure zero, therefore we cannot define $\E$ on $\Gamma$ in the usual sense unless some additional assumptions are given -- for example, regularity assumptions \ref{regul}. 
In the general case (i.e., without \ref{regul}) we overcome this difficulty by defining suitable restrictions of $\E$ and its derivatives to $\Gamma$ as elements of $\Dp(\Gamma)$ (or $\Dps(\Gamma)$).
According to the usual terminology, we will call them \textit{traces}.
Please note, that the discussion of these traces does not affect the actual interface conditions \ref{cond1f}-\ref{cond5f} ---
we only have to reformulate them in some generalized sense.

Note, that the definitions of traces standing in conditions \ref{cond1f} and \ref{cond2f} are known (see, e.g. \cite{M03}), thus we only recall them for completeness. 
\subsection{Traces arising in \ref{cond1f}}

Assume for the moment that $\E\in \mathbf{C}^1(\overline{\Rp})$. 
In this case by integrating by parts we get the following equality:
\begin{gather}\label{trace11}
\forall \Phib\in \C (\R):\ 
\intl_{\Gamma}\left(\E_+\tn\right)\cdot \Phib \d \s= 
\intl_{\Rp}(\curl \E)\cdot\Phib \d \r - \intl_{\Rp} \E\cdot(\nabla \times\Phib )\d \r.
\end{gather}
(recall that $\n=(0,0,1)^{\tt}$). One observes that the right-hand-side of Eqn.~\ref{trace11}  makes sense not only for $\E\in \mathbf{C}^1(\Rp)$, but also for 
less regular $\E$, namely $\E\in\Loc (\Rp)$ with $\curl \E\in \Loc (\Rp)$.
This suggests to introduce the following definition:
let $\E_+\in \Loc (\Rp)$,  $\curl\E_+\in\Loc (\Rp)$ (recall that generalized solutions of Eqn.~\ref{main-eq} possess  these properties); then  $\E_+\tn$ is an element of $\Dp(\Gamma)$ and its action on $\Phib\in \C(\Gamma)$  is as follows:
\begin{gather*}\label{trace1}
(\E_+\tn)[\Phib]: = 
\intl_{\Rp}(\curl \E)\cdot\widehat\Phib \d \r-\intl_{\Rp} \E\cdot(\nabla \times\widehat\Phib )\d \r,
\end{gather*}
where $\widehat\Phib$ is a function belonging to $\C(\R)$ and satisfying
$\widehat\Phib|_{\Gamma}=\Phib$.
Similarly, we define
\begin{gather*}\label{trace1+}
(\E_-\tn)[\Phib]: = 
-\intl_{\Rm}(\curl \E)\cdot\widehat\Phib \d \r+\intl_{\Rm} \E\cdot(\nabla \times\widehat\Phib )\d \r.
\end{gather*}

It is easy to see that  this definition is independent of the choice of  the function $\widehat\Phib$. Indeed, since 
\begin{gather}\label{trace12}
\intl_{\Rp} \E\cdot(\curl \widehat\Phib)\d \r\,=\, \intl_{\Rp} (\curl \E)\cdot\widehat\Phib \d\r
\end{gather}
holds for all $\widehat\Phib\in \C(\Rp)$, which is dense in $\mathcal{H}:=\{\widehat\Phib\in \C(\overline{\Rp}):\ \widehat\Phi|_\Gamma=0\}$, Eqn.~\ref{trace12} holds for $\widehat\Phib\in \mathcal{H}$, and hence for $ \widehat\Phib=\widehat\Phib_1-\widehat\Phib_2$, with $\widehat\Phib_1,\widehat\Phib_2\in \C(\overline{\Rp})$ denoting two functions satisfying $\widehat\Phib_1|_\Gamma=\widehat\Phib_2|_\Gamma$. This, obviously, gives the desired independence.

\subsection{Traces arising in \ref{cond3f}}

If $ \curl\E \in \Loc(\R)$, $\curl\curl \E\in\Loc(\Rp)$ and  $\curl\curl \E\in\Loc(\Rm)$ (due to conditions \ref{regularity} and \ref{ex1-}, generalized solutions of Eqn.~\ref{main-eq} possess these properties), we can define the traces $(\curl\E_\pm) \tn$ in the same way as the traces $\E_\pm \tn$:
\begin{gather}\label{trace3}
((\curl\E_+)\tn)[\Phib]: = 
\intl_{\Rp}(\curl\curl \E)\cdot\widehat\Phib \d \r- \intl_{\Rp} (\curl\E)\cdot(\nabla \times\widehat\Phib )\d \r,
\\
\label{trace3+}
((\curl\E_-)\tn)[\Phib]: =  
-\intl_{\Rm}(\curl\curl \E)\cdot\widehat\Phib \d \r+\intl_{\Rm} (\curl\E)\cdot(\nabla \times\widehat\Phib )\d \r.
\end{gather}
where $\widehat\Phib$ is function from $\C(\R)$ satisfying $\widehat\Phib|_\Gamma=\Phib$.

Next we clarify the meaning of 
\begin{gather}\label{G}
\mathbf{G}:=  \left(\a\curl\E_-
+\curl\gamma\curl\curl\E_- \right) \tn-
 (\Id-\n\n^{\tt})\b^z\partial_z\E_-.
\end{gather}
The idea is to define the trace for the whole quantity  $\mathbf{G}$ and not for each term separately.

Assume for the moment that $\E$ is smooth enough, namely condition~\ref{regul} holds.
Integrating by parts, one has for each $\Phib\in \C(\R)$:
\begin{multline}\label{trace31}
\intl_{\Rm }\left(\curl  \a\left(\curl\E\right)+
\suml_{j\in \{x,y,z\}}{\partial_j}\left( \b^j {\partial_j\E }\right)+
\curl\curl \g\left(\curl\curl
\E\right)\right)\cdot\Phib \d \r\\=\,
\intl_{\Rm }\left(\a\left(\curl\E\right)\cdot(\curl\Phib) 
-\suml_{j\in\{x,y,z\}}\b^j {\partial_j \E } \cdot {\partial_j \Phib }
+\g\left(\curl\curl\E\right)\cdot
(\curl\curl\Phib)\right)\d \r  \\-\,
\intl_\Gamma \mathbf{G}\cdot\Phib\d\s\,-\,
\intl_\Gamma 
\left(\left(\gamma\curl\curl\E\right)\tn\right)\cdot(\curl\Phib)\d\s+
\intl_\Gamma 
\left(\n\n^{\tt}\b^z\partial_z\E\right)\cdot \Phib\d\s.
\end{multline}
Note that the integral $\intl_{\Gamma}\mathbf{G}\cdot\Phib\d\s$ is independent of the
$z$-component of $\Phib$, since $\mathbf{G}\cdot \n=0$.
Also we observe that the integrals over $\Rm$ exist also when the regularity condition~\ref{regul} is not fulfilled, but $\E$ enjoys the properties expressed in conditions~\ref{regularity} and \ref{ex2-}.
Moreover, if we  choose the function $\Phib$ of the form given in Eqn.~\ref{Phi1}, the last two integrals in Eqn.~\ref{trace31} vanish. 

These observations suggest the following definition:
$\mathbf{G}$ is an element of $\Dp(\Gamma)$ whose action on $$\Phib=(\Phi_1(x,y),\Phi_2(x,y),\Phi_3(x,y))^\tt\in \C(\Gamma)$$ is defined by
\begin{multline}\label{trace3++}
\mathbf{G}[\Phib]:=
-\intl_{\Rm }\left(\curl  \a\left(\curl\E\right)+
\suml_{j\in \{x,y,z\}}{\partial_j}\left( \b^j {\partial_j\E }\right)+
\curl\curl  \g\left(\curl\curl
\E\right)\right)\cdot\widehat\Phib \d \r\\+\,
\intl_{\Rm }\left(\a\left(\curl\E\right)\cdot(\curl\Phib) 
-\suml_{j\in\{x,y,z\}}\b^j {\partial_j \E } \cdot {\partial_j \Phib }
+\g\left(\curl\curl\E\right)\cdot
(\curl\curl\widehat\Phib)\right)\d \r,
\end{multline}
where $\widehat\Phib(\r)=(\Phi_1(x,y)\eta(z),\,\Phi_2(x,y)\eta(z),\,0)^\tt$,
with $\eta\in \Cs(\mathbb{R})$ satisfying $\eta|_{\{|z|<\delta\}}=1$  for some $\delta>0$. It is easy to see that this definition is independent of the function $\eta$ -- the proof is the same as the one for $\E_+\tn$.

\subsection{Traces arising in \ref{cond4f}}

To define the trace $(\gamma\curl\curl \E_-)\tn$ we use an analogous  
idea. This time we plug into Eqn.~\ref{trace31} the function $\Phib$ of the form given in Eqn.~\ref{Phi2}. Recall that for this $\Phib$ one has $\Phib|_\Gamma=0$ (and, thus, the first and the last integrals over $\Gamma$ in \ref{trace31} vanish), while $ \curl\Phib|_{\Gamma}=(\Phi_1,\,\Phi_2,\ 0)$. 

Taking these observations into account we define  $(\gamma\curl\curl \E_-)\tn$ as an element of $\Dp(\Gamma)$ whose action on $\Phib=(\Phi_1(x,y),\Phi_2(x,y),\Phi_3(x,y))^\tt\in \C(\Gamma)$
is defined by the right-hand-side of \ref{trace3++}
with $\widehat\Phib(\r)=(\Phi_2(x,y)z\eta(z),\,-\Phi_1(x,y)z\eta(z),\,0)^\tt$ (the functions $\eta$ is as above).

\subsection{Traces arising in \ref{cond5f}}

Let $\E$ satisfy the properties given in the conditions \ref{regularity} and \ref{ex2-}. 
Then we define the trace $\left(\b^z\partial_z\E_-\right)\cdot \n$ as an element of $\Dps(\Gamma)$ whose action on $\Psi\in \Cs(\Gamma)$ is defined by 
\begin{multline*}
(\left(\b^z\partial_z\E_-\right)\cdot \n)[\Psi]:=
\intl_{\Rm }\left(\curl  \a\left(\curl\E\right)+
\suml_{j\in \{x,y,z\}}{\partial_j}\left( \b^j {\partial_j\E }\right)+
\curl\curl  \g\left(\curl\curl
\E\right)\right)\cdot\widehat\Phib \d \r\\-\,
\intl_{\Rm }\left(\a\left(\curl\E\right)\cdot(\curl\Phib) 
-\suml_{j\in\{x,y,z\}}\b^j {\partial_j \E } \cdot {\partial_j \Phib }
+\g\left(\curl\curl\E\right)\cdot
(\curl\curl\widehat\Phib)\right)\d \r\\+\,\left((\gamma\curl\curl \E_-)\tn\right)[\curl\widehat\Phib|_\Gamma],
\end{multline*}
where $\widehat\Phib(\r)=(0,0,\Psi\eta)^\tt$ (the functions $\eta$ is as above). 

The reason for such a definition is as follows. 
If we plug into Eqn.~\ref{trace31} the function $\Phib(\r)=(0,0,\Psi\eta)^\tt$, the first integral over $\Gamma$ vanishes, and the integral 
$\intl_\Gamma 
\left(\n\n^{\tt}\b^z\partial_z\E\right)\cdot \Phib\d\s$ becomes
$\intl_\Gamma 
(\left(\b^z\partial_z\E\right)\cdot \n) \Psi\d\s$. Moreover,
the integrals over $\Rm$ exist not only for smooth $\E$, but also when the properties expressed in conditions \ref{regularity} and \ref{ex2-} hold. Finally, the integral $\intl_\Gamma 
\left(\left(\gamma\curl\curl\E\right)\tn\right)\cdot(\curl\Phib)\d\s$ can be interpreted as the action of $\left(\gamma\curl\curl\E\right)\tn$ on  
$\curl\Phib|_\Gamma$.

\subsection{Traces  arising in \ref{cond2f}}

To define the trace  $(\curl \E_+)\cdot \n$ we use the equality
\begin{gather}
\label{trace21}
\intl_{\Gamma}\left((\curl \E_+)\cdot \n\right) \Psi  \d \s= -
\intl_{\Rp}(\curl \E)\cdot(\nabla\Psi )\d \r,
\end{gather}
which hold for all $\Psi\in\Cs(\R)$ provided $\E\in \C(\overline{\Rp})$.
The integral  in the right-hand side of Eqn.~\ref{trace21} exists even for less regular $\E$. Namely, one needs only $\curl\E\in \Loc(\Rp)$ (which is again fulfilled for generalized solutions of Eqn.~\ref{main-eq}). This suggests to define  
$(\curl \E_+)\cdot \n$  as an element of $\Dps(\Gamma)$ acting on $\Psi\in \Cs(\Gamma)$ as follows:
\begin{gather*}
((\curl \E_+)\cdot \n)[\Psi]:= -
\intl_{\Rp}(\curl \E)\cdot(\nabla\widehat\Psi )\d \r,
\end{gather*}
where $\widehat\Psi$ is function from $\Cs(\R)$ satisfying $\widehat\Psi|_\Gamma=\Psi$. Similarly, we define
\begin{gather*}
((\curl \E_-)\cdot \n)[\Psi]:= 
\intl_{\Rm}(\curl \E)\cdot(\nabla\widehat\Psi )\d \r.
\end{gather*}

\subsection{Traces  arising in \ref{cond-add}}

Let $\E\in\Loc(\Rp)$ and $\div\E=0$ in $\Dps(\Rp)$ (these conditions are fulfilled for 
generalized solutions of Eqn.~\ref{main-eq}).
Similarly to $(\curl\E_+)\cdot \n$ we define
the trace $\E_+\cdot \n$ as an element of $\Dps(\Rp)$ acting on $\Psi\in \Cs(\Rp)$ as follows:
\begin{gather*}
(\E_+\cdot \n)[\Psi]:=-\intl_{\Rp}\E\cdot\nabla\widehat\Psi\d\r,
\end{gather*}
where $\widehat\Psi$ is function from $\Cs(\R)$ satisfying $\widehat\Psi|_\Gamma=\Psi$.

And, finally, we examine the trace
$$W:=\left(\eps\E_- 
+\suml_{j\in\{x,y,z\}}\partial_j\left(\b^j {\partial_j \E_- }\right) \right)\cdot \n+\nabla_\Gamma\cdot\left(\b^z\partial_z\E_z\right).$$
Assuming that $\E$ satisfies \ref{regul} one has for each $\Psi\in\Cs(\Rm)$:
\begin{multline}\label{trace-add1}
\intl_\Gamma W\Psi\d\s=
\intl_\Gamma \left((\beta^z\partial_z\E)\cdot \n\right)\partial_z\Psi\d\r\\
+\intl_{\Rm}\div\left(\eps\E+\suml_{j\in\{x,y,z\}}
\partial_j(\beta^j\partial_j\E)\right)\Psi\d\r+
\intl_{\Rm}\left(\eps\E\cdot\nabla\Psi-\suml_{j\in\{x,y,z\}}
\beta^j\partial_j\E\cdot\partial_j\nabla\Psi\right)\d\r.
\end{multline}
Equality \ref{trace-add1} suggests the following definition: let
$\E$ satisfies the regularity condition given in expressions \ref{regularity}, \ref{ex2-}, and \ref{cond-add4} holds; we define 
the trace $W$ as an element of $\Dps(\Gamma)$ acting on $\Psi\in \Cs(\Gamma)$ as follows:
\begin{gather*}
W[\Psi]:=
\left((\beta_z\partial_z\E)\cdot \n\right)[\partial_z\widehat\Psi]\,+
\intl_{\Rm}\left(\eps\E\cdot\nabla\widehat\Psi-\suml_{j\in\{x,y,z\}}
\beta^j\partial_j\E\cdot\partial_j\nabla\widehat\Psi\right)\d\r,
\end{gather*}
where $\widehat\Psi$ is a function from $\Cs(\R)$ satisfying $\widehat\Psi|_\Gamma=\Psi$.

\subsection{Interface conditions for non-smooth generalized solutions}

Now, assume that $\E$ is a generalized solution of Eqn.~\ref{main-eq}. This means that $\E$ satisfies expressions \ref{regularity}, \ref{main-inteq} and, as a result, conditions \ref{ex1-}-\ref{ex2}, \ref{cond-add4}-\ref{cond-add5} hold, but the fulfilment of \ref{regul} is not guaranteed, in general. Nevertheless, it is easy to see that the interface conditions \ref{cond1f}-\ref{cond5f}, \ref{cond2f}-\ref{cond-add} still hold with traces being  understood according to the above definitions --- as   elements of $\Dp(\Gamma)$ or $\Dps(\Gamma)$.

Let us check this, for example,  for condition \ref{cond3f}.
For the function $\Phib=(\Phi_1,\Phi_2,\Phi_3)\in\C(\Gamma)$ we define the function $\widehat\Phib$ by the formula \ref{Phi1}. 
One has (below we use the notation $\mathbf{G}$, which is defined in \ref{G}): 
\begin{multline*}
0\overset{\ref{ex1-}-\ref{ex2}}{=}\intl_{\Rp}\left((\curl\curl\E) -
{\omega^2 \over c^2} 
\E\right)\cdot\widehat\Phib\d\r+ \intl_{\Rm}\left((\curl\curl\E)  -
{\omega^2 \over c^2} \Big(\eps \E\right.
\\
\left.+\a\left(\curl\curl\E\right) 
+\suml_{j\in\{x,y,z\}}\partial_j\left(\b^j {\partial_j \E }\right)  
+\curl\curl\g\left(\curl\curl\E\right)\Big)\right)\cdot \widehat\Phib\d \r \\\overset{\ref{trace3},\ref{trace3+},\ref{trace3++}}{=}
\left((\curl\E_+)\tn\right)[\Phib]-\left((\curl\E_-)\tn\right)[\Phib]
+{\omega^2\over c^2}\mathbf{G}[\Phib]+\intl_{\R} (\curl\E)\cdot (\curl\widehat\Phib)\d \r\\\,-
{\omega^2 \over c^2} 
\intl_{\Rm} 
\Big(\eps \E\cdot\widehat\Phib
+\a\left(\curl\E\right)\cdot(\curl\widehat\Phib) 
-\suml_{j\in\{x,y,z\}}\b^j {\partial_j \E } \cdot {\partial_j \Phib }
+\g\left(\curl\curl\E\right)\cdot
(\curl\curl\widehat\Phib)\Big)\d \r\\\overset{\ref{main-inteq}}{=}
\left((\curl\E_+)\tn\right)[\Phib]-\left((\curl\E_-)\tn\right)[\Phib]
+{\omega^2\over c^2}\mathbf{G}[\Phib]
\end{multline*}

As we have shown in this section, the interface conditions as such are not affected by these considerations.

\section{Conclusions}
In conclusions, we have been considering light propagation in metamaterials that are described at the effective level with nonlocal constitutive relations. Nonlocal laws have also been considered in the past but their discussion was usually linked to a specific metamaterial geometry. Also, the interface problem has been solved by introducing additional boundary conditions on phenomenological ground. Our contribution distinguishes itself that the nonlocal constitutive relation we consider is a firm extrapolation from the previous understanding that applies to arbitrary metamaterials. The dispersion relation of such metamaterials has been already discussed in the past; and here we have equipped these dispersion relations with the necessary interface conditions.

These interface conditions are derived from first principles. They rely on the evaluation of a generalized formulation of Maxwell's equations in a small volume entailing the interface. We have been deriving four main interface conditions and also formulated two alternative conditions. They do not introduce further information but basically look simpler. This might be occasionally beneficial. For a single interface, the consideration of all interface conditions is necessary. This might sound surprising but the nonlocal metamaterial sustains multiple plane waves as eigenmodes at each given frequency. This is in contrast to a local medium where only a single plane wave is supported. 

Besides the actual interface conditions, we also derived explicit expressions for the Fresnel equations that can predict  reflection and transmission from a slab upon illuminating it with a plane wave. We discuss both TE and TM polarization. The Fresnel equations are documented in a convenient way and are expressed in matrix notation.

{With this work}, further research endeavours in the context of metamaterials are immediately possible where the physics of such nonlocal metamaterials can be explored. It starts by developing suitable retrieval procedures and the actual quantification of the nonlocality of the metamaterials. It can be extended by analysing basic optical phenomena in the presence of metamaterials with a strong nonlocality and studying potential applications that rely on such nonlocal metamaterials. Also, the development of suitable numerical tools to explore light propagation in nonlocal metamaterials is an important and timely issue. Finally, based on the general formalism, other kinds of nonlocal constitutive relations can be explored. 

\section*{Appendix}

In this appendix, we present a short overview on the concept of generalized functions. For more details we refer, e.g. to Ref.~\onlinecite{F82}.

\subsection*{Generalized functions}

Let $\Omega$ be an open set in $\mathbb{R}^n$ (in the paper, $\Omega$ is either $\R$, $\Rm$, $\Rp$ or $\Gamma$). Recall that $\Cs(\Omega)$ denotes  the space of infinitely often differentiable functions $\Psi: \Omega \to \mathbb{C}$ with compact support  $\supp(\Psi):=\overline{\{x:\ \Psi(x)\not=0 \}}\subset\Omega$. Note, that in  textbooks concerned with generalized functions this space is usually denoted by $\mathsf{D}(\Omega)$, but we prefer the notation $\Cs(\Omega)$ in order to  avoid possible confusions with the electric displacement field $\D$.

The space $\Cs(\Omega)$ can be equipped with a concept of convergence: a sequence $\{\Psi_k\}_{k\in\mathbb{N}}$ in $\Cs(\Omega)$ is said to converge to 
$\Psi\in \Cs(\Omega)$ ($\Psi_k\to\Psi$ in $\Cs(\Omega)$) if
there is a compact set $K \subset \Omega$ containing the supports of all $\Psi_k$, and
for each multi-index $\a$ the sequence of partial derivatives $\left\{\partial_\a\Psi_k\right\}$
converges uniformly to $\partial_\a\Psi$.

A \textbf{generalized function} (or a \textbf{distribution})
 on $\Omega$ is a continuous linear functional $F : \Cs(\Omega) \to \mathbb{C}$, i.e.,  $F$ assigns to each function $\Psi\in \Cs(\Omega)$ a complex number $F[\Psi]$ so that
\begin{itemize}
\item    $F(c_{1}\Psi _{1}+c_{2}\Psi _{2})=c_{1}F(\Psi _{1})+c_{2}F(\Psi _{2})$ for all $\Psi_1,\Psi_2\in\Cs(\Omega)$, $c_1,c_2\in\mathbb{C}$.

\item $\liml_{k\to\infty}F[\Psi_k]= F[\Psi]$ provided 
$\Psi_k \to\Psi$ in $\Cs(\Omega)$.

\end{itemize}
The space of generalized functions on $\Omega$ is denoted by $\Dps(\Omega)$.

\subsection*{Examples}
Recall, that $\mathbf{L}^1_{\rm loc}(\Omega)$ is the set of locally integrable functions (i.e., the functions being Lebesgue integrable over every compact subset of $\Omega$).
Any function $f\in \mathbf{L}^1_{\rm loc}(\Omega)$ yields a generalized function $F_f$ acting on $\Psi\in \Cs(\Omega)$ by
\begin{gather}\label{reg-distr}
F_f[\Psi]:=\intl_\Omega f(x)\Psi(x)\d x.
\end{gather}

If $F\in \Dps(\Omega)$ is given by \ref{reg-distr} for some  $f\in \mathbf{L}^1_{\rm loc}(\Omega)$, then $F$ is called a \textbf{regular} generalized function.

Let $f,g\in \mathbf{L}^1_{\rm loc}(\Omega)$. Then the associated regular generalized functions $F_f$ and $F_g$  coincide (i.e. $F_f[\Psi]=F_g[\Psi]$, $\forall\Psi$) if and only if $f$ and $g$ are the same elements of $\mathbf{L}^1_{\rm loc}(\Omega)$ (i.e., they are are equal almost everywhere).

In what follows we will use the same notation for a locally integrable function and for the associated regular generalized function.

$F\in\Dps(\Omega)$ is called \textbf{singular} if it is not a regular generalized function. An example of a singular generalized function is a \textit{Dirac delta-function} $\delta_{x_0}\in \Dps(\Omega)$ supported at $x_0\in\Omega$: its acts on $\Psi\in \Cs(\Omega)$ by $\delta_{x_0}[\Psi]:=\Psi(x_0).$

\subsection*{Operations on generalized function}

\subsubsection*{Linear combination}
Let $F_k\in\Dps(\Omega)$ and $c_k\in\mathbb{C}$, $k=1,\dots,N$. Then the generalized function
$F=\sum_{k=1}^N c_k F_k$ acts on  $\Psi\in \Cs(\Omega)$ by
$$F[\Psi]:=\suml_{k=1}^N c_k F_k[\Psi].$$

\subsubsection*{Differentiation}

Let $F\in\Cs(\Omega)$ and $\Psi\in \Cs(\Omega)$. 
Let $\a=(\a_1,\dots\a_n)$ be a multindex and $\ds\partial_\a:={\partial^{\a_1}\over\partial x_1^{\a_1}}{\partial^{\a_2}\over\partial x_2^{\a_2}}\dots {\partial^{\a_n}\over\partial x_n^{\a_n}}$ be the associated partial differential operator. Integration by parts gives
\begin{gather}\label{int-by-parts}
\intl_{\Omega}(\partial_\a F)\,\Psi\d x=(-1)^{|\a|}\intl_{\Omega}F\,(\partial_\a \Psi)\d x,
\end{gather}
where $|\a|=\a_1+\a_2+\dots+\a_n$.

Now, let $F\in\Dps(\Omega)$. Equality \ref{int-by-parts} suggests to define
$\partial_{\a}F$ as an element of $\Dps(\Omega)$ acting on $\Psi\in \Cs(\Omega)$ by
$$(\partial_{\a}F)[\Psi]:=(-1)^{|\a|}F[\partial_{\a}\Psi].$$

As we see, in contrast to   ``classical'' functions, generalized functions can be differentiated infinitely often. Of course, if $F$ is a regular  generalized function, $\partial_\a F$ need not be regular, in general. For example, let $H(x)$ be the Heaviside step function on $\mathbb{R}$ being equal to $1$ as $x>x_0$ and equal to $0$ as $x\leq x_0$. $H(x)$ is not differentiable on $\mathbb{R}$ due to the discontinuity at $x_0$.
But, since $H\in \mathbf{L}^1_{\rm loc}(\mathbb{R})$ we can regard it as a generalized function. 
It is easy to see that the generalized derivative of $H$ is $\delta_{x_0}$.

\subsubsection{Multiplication by a smooth function}

Let $f$ be an infinitely differentiable function on $\Omega$ and $F\in\Dps(\Omega)$. 
We define the product $fF$ as an element of   $\Dps(\Omega)$ acting on $\Psi\in \Cs(\Omega)$ by
$$(fF)[\Psi]:=F[f\Psi].$$
This definition is motivated by the commutativity of the product of scalar functions, namely
$$\intl_\Omega f(x) F(x)\, \Psi(x)\d x=\intl_\Omega F(x)\, f(x)\Psi(x) \d x$$
provided $F\in\mathbf{L}^1_{\rm loc}(\Omega)$.

It is clear that the above definition fails if $f$ is not smooth since the product $f\Psi$ is no longer in $\Cs(\Omega)$. On the other hand if $F$ is a regular generalized function and $f$ is locally bounded then the product is well-defined via $(fF)[\Psi]:=\intl_\Omega f(x) F(x)\, \Psi(x)\d x.$

\subsection*{Generalized vector-functions}

Recall, that $\C(\Omega)$ stands for  the space of infinitely differentiable vector-functions $\Phib: \Omega \to \mathbb{C}^3$ with compact support in $\Omega$. Then we define $\Dp(\Omega)$  as the space of continuous linear functionals acting  on $\C(\Omega)$.

The definition of derivatives for $\Dp(\Omega)$ is similar to the corresponding definition for $\Dps(\Omega)$ -- it mimics partial integration.   In particular, for $\mathbf{F}\in \Dp(\Omega)$ one has $\curl \mathbf{F}\in \Dp(\Omega)$,  $\div\mathbf{F}\in \Dps(\Omega)$, by the definitions
\begin{gather*}
(\curl \mathbf{F})[\Phib]:= \mathbf{F}[\curl \Phib],\quad
(\div \mathbf{F})[\Psi]:= -\mathbf{F}[\nabla\Psi].
\end{gather*}

\section*{Acknowledgement}
We gratefully acknowledge financial support by the Deutsche Forschungsgemeinschaft (DFG) through CRC 1173. K.M. also acknowledges support from the Karlsruhe School of Optics and Photonics (KSOP). C.S. acknowledges the support of the Klaus Tschira Stiftung.

\newpage\small


\end{document}